# Visual Search for Galaxies near the Northern Crossing of the Supergalactic plane by the Milky Way.

G. K. T. Hau[1,3], H. C. Ferguson[2,1], O. Lahav[1], and D. Lynden-Bell[1]
[1] *Institute of Astronomy, Madingley Road., Cambridge, CB3 0HA*
[2] *Space Telescope Science Institute, 3700 San Martin Drive, Baltimore, MD 21218, USA*
[3] *E-mail address: gkth@mail.ast.cam.ac.uk*

**ABSTRACT**
We have visually examined twelve Palomar red Plates for galaxies at low Galactic latitude $b$, where the Supergalactic Plane (SGP) is crossed by the Galactic Plane (GP), at Galactic longitude $l \sim 135°$. The catalogue consists of 2575 galaxy candidates, of which 462 have major axis diameters $d \geq 0.8'$ (uncorrected for extinction). Galaxy candidates can be identified down to $|b| \approx 0°$. One of our galaxy candidates (J24 = Dwingeloo 1) has recently been discovered independently in 21cm by Kraan-Korteweg et al. (1994) as a nearby galaxy. Comparisons with the structures seen in the IRAS and UGC catalogues are made. We compare the success rate of identifying galaxies using the IRAS Point Source Catalogue under different colour selection criteria. The criteria that require both the 60 $\mu m$ and 100 $\mu m$ fluxes to be of high quality, have the highest probability of selecting a galaxy (with $d \geq 0.6'$), but at the expense of selecting a smaller number of galaxies in total.

**Key words:** Cosmology: large-scale structure of the Universe – Galaxies: individual: Dwingeloo 1 – Galaxies: Local Group – ISM: dust, extinction

## 1 INTRODUCTION

A major obstacle to the study of large scale structure is the obscuration or confusion of $\sim 10 - 20\%$ of the extragalactic sky by the Galactic Plane, the so-called Zone of Avoidance (ZOA). Important structures such as the Supergalactic plane, the Hydra-Centaurus region, the Perseus-Pisces supercluster and the Puppis cluster all disappear behind the ZOA. The 'missing galaxies' may have important implications for the origin of tidal torques exerted by neighbouring galaxies, for the internal dynamics of the Local Group, for the origin of its motion relative to the Microwave Background, and for the connectivity of large scale structure. Methods for unveiling the ZOA include visual examination of plates, HI observations at 21 cm, infrared and X-ray observations, as well as statistical interpolation methods. For review see Kraan-Korteweg & Woudt (1994), Lahav (1994) and Weinberger et al. (1994) and references therein. The importance of probing the ZOA is indicated by the fact that several bright galaxies lie behind the Galactic Plane (e.g. Cen A, IC342, Maffei I and II and N4945). Since galaxies are clustered, it would not be too surprising to discover more galaxies in the neighbourhood of these known galaxies.

Several groups have demonstrated that useful information can be obtained by visually examining photographic plates (Kraan-Korteweg & Woudt 1994, Saitō et al. 1990, Yamada et al. 1993). Here we present a catalogue of galaxy candidates visually identified on 12 Palomar red plates. The region scanned is $l \approx 135°$, $-20 \lesssim b \lesssim 20°$, where the Supergalactic Plane is crossed by the Galactic Plane. This region is a promising place for galaxy search, as it lies close to the Perseus Pisces Supercluster, and includes many nearby galaxies such as Maffei I, Maffei II and IC342. An encouraging support for the reliability of our list of candidates has been provided recently by an independent observation in the 21 cm line of one of the galaxies in our list J24 = Dwingeloo 1 (Kraan-Korteweg et al. 1994).

The outline of the paper is as follows. In §2 we describe the search strategy and the catalogue, which is analyzed in §3. Comparison with other catalogues is provided in §4. In §5 the digitized images of several galaxy candidates are presented, followed by discussion in §6.

## 2 THE SEARCH

### 2.1 The search criterion

A total of 12 Palomar red (E) plates were scanned, covering an area of $\sim 444$ $deg^2$; Table 1 displays the list of plates and the order they were scanned. The red plates were preferred to the blue plates after a trial scan. The plates were examined to include all galaxies that have diameters $d \geq 0.6'$ (not corrected for extinction) in order to give a high confidence of completeness at $d = 0.8'$. Many galaxies with diameters down to $d \sim 0.3'$ were also included in the catalogue to ensure completeness.

**Table 1.** Scanning order of the plates, the plate numbers, the coordinates of the plate centres and the private plate label.

| Scan order | Plate | $\alpha_{1950}$ | $\delta_{1950}$ | Label |
|---|---|---|---|---|
| 1 | 1241 | 2 44 04 | 66 24 02 | A |
| 2 | 1230 | 2 24 50 | 72 25 40 | B |
| 3 | 597 | 2 18 59 | 60 26 02 | C |
| 4 | 1245 | 2 00 19 | 54 27 41 | D |
| 5 | 907 | 2 22 14 | 48 26 06 | E |
| 6 | 931 | 2 05 52 | 42 27 09 | F |
| 7 | 1226 | 3 16 30 | 78 21 46 | G |
| 8 | 1617 | 2 38 43 | 54 24 42 | H |
| 9 | 865 | 3 34 19 | 72 17 56 | I |
| 10 | 968 | 3 03 27 | 60 22 29 | J |
| 11 | 845 | 2 56 30 | 48 22 19 | K |
| 12 | 973 | 3 36 40 | 66 18 48 | L |

Galaxy diameters can be corrected for extinction in the blue ($A_B$) by the empirical relation (Cameron 1990)

$$d_{true} = 10^{\alpha A_B^\beta} d_{obs}, \quad (1)$$

with the coefficients $\alpha = 0.10$, $\beta = 1.7$ for spirals, and $\alpha = 0.13$, $\beta = 1.3$ for ellipticals and S0's. If colour gradients are small over the galaxy, the above relation can be used in other wavebands, with $A_R \approx 0.5 A_B$ and $A_V \approx 0.75 A_B$ (e.g. Johnson 1965). As an example, a spiral hidden under 6 magnitudes of blue extinction, typical for $|b| \approx 0$, will have a blue diameter $\sim 1\%$ its true value, and a red diameter $\sim 25\%$ its true value.

### 2.2 Scanning Procedure

The visual examination was done by one of us (GKTH) in 1991. As with any other visual surveys, it is inevitable that some systematics and variation in sensitivity both across and between the plates will be introduced. Before the search program began, a plate was scanned for training so as to reduce the effect of increased efficiency as the search progressed. The plates were accessed in the order given in column 1 of Table 1, and individual plates were scanned in a zig-zag pattern down the plate to avoid any gradual variation of sensitivity across the plate. A traveling microscope at 6× magnification was used. Before each scan the plate was viewed with the naked eye to look for unusually faint and diffuse features that may be missed by looking through the microscope. Typically a plate takes about 4 hours to scan. For each galaxy, the position, major-axis diameter, morphological type are recorded, as well as any comment. The major-axis diameter is defined as the maximum extent of the galaxy as seen by the eye (usually through the microscope), and are specified in increments of $0.05'$. The rms scatter of the diameters for the 126 galaxies in the plate overlaps is found to be $0.2'$; the overall rms scatter is expected to be smaller than this value, as the diameters are usually difficult to measure near the plate edges. The positions are measured using a coradograph. By analyzing the positional difference of the galaxies in the plate overlaps, the rms error for the positions is found to be about $15''$. When compiling the final catalogue, the positions and diameters of the galaxies in the plate overlaps are replaced by the averaged quantities.

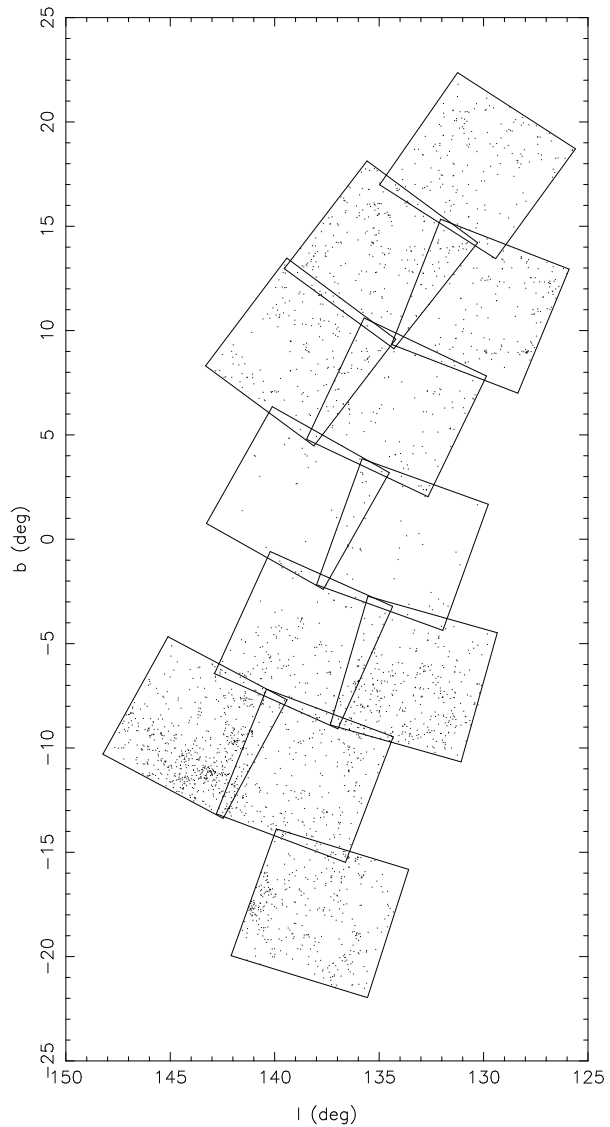

**Figure 1.** All 2575 galaxies in our sample. The plate edges are also shown

### 2.3 The projected distribution of the sample

A total of 2575 galaxy candidates are found in the search region. The corresponding number of galaxies to diameter limits of $0.6', 0.8', 1.0'$ are 936, 462 and 274 respectively. The complete catalogue is plotted in Fig. 1 together with the plate boundaries, whilst the galaxies with $d \geq 0.8'$ are plotted in Fig. 2. In Fig. 3 the contours of Galactic HI from Stark et al. (1992) are plotted. From Diplas & Savage (1994) the red extinction $A_R$ can be related to the HI column density by

$$A_R \approx \frac{2}{3} A_V \approx 2E(B-V) \approx \frac{N_{\rm HI}}{2.465 \times 10^{21} {\rm cm}^{-2}} \text{ mag}, \quad (2)$$

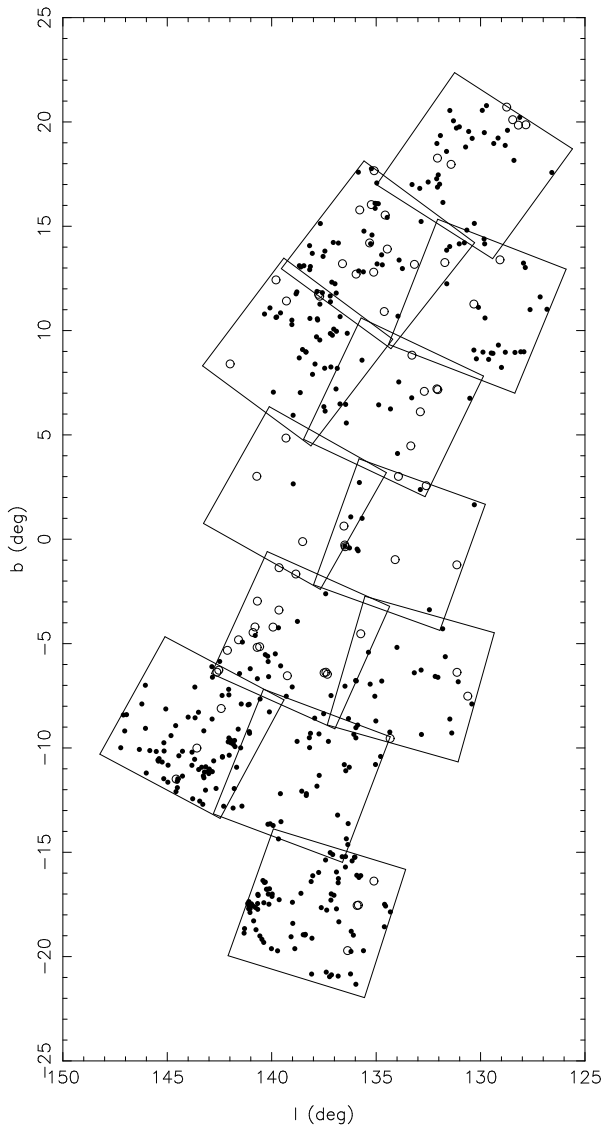

**Figure 2.** The 462 galaxies with $d \geq 0.8'$. The open circles (69) are entries that are potentially noise, whilst the filled circles (393) are the the entries that do not have a 'N' noise flag.

**Figure 3.** Galactic HI column density from Stark et al. The contours are in increments of $5 \times 10^{20}$ cm$^{-2}$; the top contour has value $8 \times 10^{21}$ cm$^{-2}$.

although this relation may not hold tightly in the galactic plane. Galaxies can be seen down to $b = 0°$, but there is a rapid drop in detection rate for $-6° \lesssim b \lesssim 8°$. This corresponds to an HI column density of $\gtrsim 3.5 \times 10^{21}$cm$^{-2}$.

The survey area is roughly equal for the Galactic north (230 deg$^2$) and the south (215 deg$^2$). However, over the entire catalogue there are more galaxies (with diameters not corrected for extinction) detected for $b < 0°$ (1742) than for $b > 0°$ (833). At the diameter limit of $0.8'$, the corresponding numbers are 279 and 183 respectively. The Galactic north-south asymmetry is more likely to be due to clustering than due to Galactic extinction. This is supported by the Galactic HI map of Stark et al. (1992) shown here in Fig. 3, indicating north-south symmetry about the Galactic Plane. Over-dense regions can be seen at $(l, b) \approx (141°, -18°)$ and $(143°, -12°)$ on Fig. 2, and is best seen in Fig. 6. Figure 4 shows our survey together with UGC, ESO and MCG galaxies (e.g. Lahav 1987). Although the four surveys are not matched and not corrected for extinction, the visual impression is that we see the connectivity of the Supergalactic Plane across the Galactic Plane.

In order to investigate the likely effect of confusion due to Galactic sources and misidentification in our sample, in Fig. 2 we mark with open circles the $d \geq 0.8'$ galaxies with an occurance of a 'N' (for potentially 'noise') in the 'Morphological Type' column, explained in §2.4. There are 69 such entries, about 2/5 of these are located near the plate edges. Although the fraction of galaxies that are potentially noise increases with decreasing $|b|$, the number density of these entries remains roughly uniform over the whole surveyed area. In particular, galaxies can still be identified behind the galactic plane even when the 'noisy' entries are discarded.

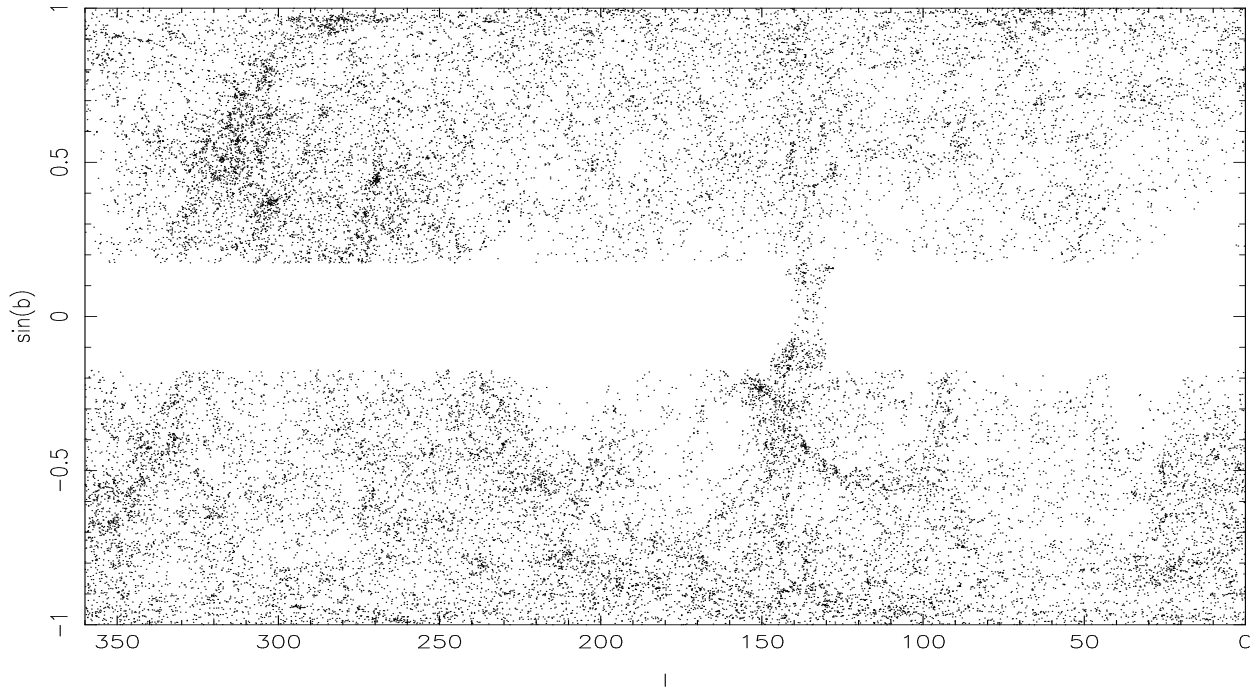

**Figure 4.** The location of our survey with respect to the rest of sky. The abscissa is galactic longitude $l$ and the ordinate is $sin(b)$. For $|b| > 10$ the UGC and ESO galaxies are plotted to the (raw) diameter limit of $1'$. The strip missed by UGC and ESO is supplemented by the galaxies from MCG. For $|b| < 10'$, our visually identified galaxies at a diameter limit of $0.6'$ are plotted.

### 2.4 The Catalogue

The 274 galaxies with $d \geq 1.0'$ are listed in Tables 2a and 2b. A separate list of all the 2575 galaxies in the survey are given on the enclosed microfiche (Tables 6a & 6b). The catalogues can be supplied in electronic form on request from GKTH. The columns in Tables 2a and 6a are:

**Column 1** The galaxy name in the notation of G$lll.ll+/-b.bb$ as suggested by Weinberger et al. (1995), where $lll.ll$ and $b.bb$ are the galactic longitude and latitude respectively.

**Columns 2,3** The Right Ascension and Declination (epoch 1950) of the galaxy centre. For objects that are spotted more than once in the plate overlaps, the mean position is given. For galaxies in tight groups, one single entry is given and the position of the centre of the distribution is listed.

**Column 4** Our identification label. The character denotes the plate label (as listed in Table 1) and the number denotes the position in the list of galaxies spotted on that plate. Hence an identification 'A25' refers to the 25th galaxy spotted on POSS plate E1241. Galaxies that are identified more than once in the plate overlaps will have composite labels, hence a label 'A37L59' means that the galaxy is spotted on POSS plates E1241 and E973.

**Column 5** Morphological type as judged by the eye, crudely classified under S, SB, S0, E, DE, I, U, * and N for spiral, barred-spiral, S0, elliptical, dwarf elliptical, irregular, unknown, star or noise. In regions of high extinction and galactic contamination, all likely morphological types are listed, ordered according to their likelihood. For example, a morphological type of 'S/N' means that the galaxy is likely to be a spiral, but with an alternative classification of noise. Note that spiral and elliptical classifications never appear together; in such a case, the galaxy type is likely to be classified as 'U' (for unknown).

**Column 6** Major axis diameter, in arc minutes. For galaxies in tight groups (e.g. mergers), such that they are taken as one entry, the overall diameter is listed.

**Column 7** IRAS 60 $\mu$m flux.

**Column 8** UGC blue diameter (Nilson 1973).

**Column 9** UGC number.

**Column 10** B magnitude as listed by Huchra ZCAT.

**Columns 11, 12, 13** Heliocentric velocity in km s$^{-1}$ listed in the Huchra's ZCAT that was available to us in 1991, velocity error, and velocity flag (1 =velocity is specified in ZCAT, $-1$ = velocity is measured but not made available to the public).

**Column 14** Name as listed in column 1 of ZCAT.

The comments on individual candidate are listed separately in Tables 2b and 6b.

### 3 STATISTICS

#### 3.1 Number counts

The cumulative number count for the whole sample, normalized to number per square degrees, is plotted in Fig.5 (thick solid line). The number count follows well a power law of slope $-2.2$ for $d \gtrsim 0.5'$, and deviates from the power law for smaller diameters due to incompleteness. There may be a slight excess for $d \gtrsim 2.3'$. If galaxies are distributed uniformly in space, the slope of the graph should be $-3$;

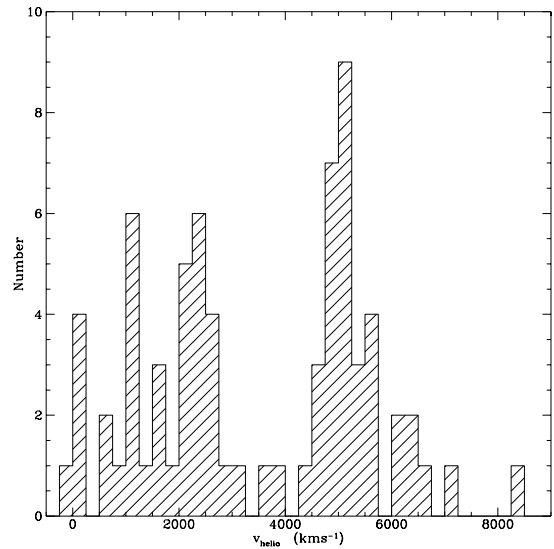

**Figure 7.** The heliocentric redshift distribution of our 70 $d \geq 1'$ galaxies that have listed redshifts in Huchra's ZCAT. The recently discovered Dwingeloo 1 (our candidate J24) is included.

**Figure 5.** The cumulative number counts, normalized to number per square degrees, for the whole catalogue, $b < -10°$, $-10° < b < 0°$, $0° < b < 10°$ and $b > 10°$. The number counts in the Galactic south are shifted up by a factor of 10.

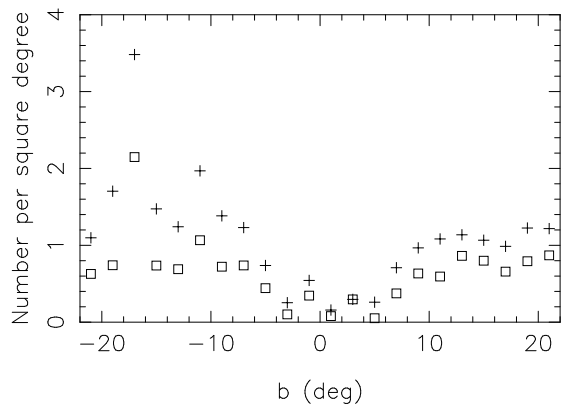

**Figure 6.** The number density (per square degrees) of eyeballed galaxies. The crosses are for $d \geq 0.8'$, whilst the squares are for $d \geq 1.0'$.

however, since the search region is known to contain various nearby structures, and is under heavy obscuration, a slope which differs from $-3$ is not unexpected. Whilst we are confident that our sample is complete down to the raw diameter of $0.8'$, the plot suggests further that our sample is highly complete down to raw diameter of $0.6'$, as indicated by the turnoff at $d \sim 0.5'$.

In order to investigate further the dependency of the number counts with galactic latitudes, the cumulative number counts for different $b$ slices are plotted in Fig. 5. For $-10° < b < 0°$, the number count follows well a slope of $-2.5$ for $0.5' < d < 1.4'$, but shows an excess of galaxies for $d \gtrsim 1.4'$, and exceeding the number counts of $b < -10°$ at $d \gtrsim 1.8'$. For both $0° < b < 10°$ and $b > 10°$, the number counts are shallower than average. The reason why the slope of number counts varies with different galactic latitude slices may be due to different contribution from foreground and background objects to our sample (see §3.2 and Fig. 8). Although Fig. 8 is incomplete and possibly biased, the domination by objects with $cz > 4000\,\mathrm{km\,s^{-1}}$ in the lower part of Fig. 8 may explain why the number count for $b < -10°$ is steeper than the rest. The shallower number counts for $b > 0°$ as well as the excess of galaxies larger than $d \sim 1.4'$ for $-10° < b < 0°$ may be due a higher contribution of more nearby galaxies to our sample. Alternatively, the excess of large galaxies in the number counts could be caused by patchy extinction such that in 'holes' of low extinction the galaxy would appear larger than expected.

### 3.2 The redshift distribution of the sample

Down to diameter $d = 1'$ there are 101 galaxies with measured redshifts from a compilation by Huchra (as was available to us in 1991) within a search radius of 2 arc minutes, of which 70 are published. We supplement the redshifts of these 70 galaxies with that of the recently discovered Dwingeloo 1 (which is our galaxy candidate J24) at heliocentric redshift of $\sim 110\,\mathrm{km\,s^{-1}}$(Kraan-Korteweg et al. 1994). The heliocentric redshift distribution are plotted in Fig. 7. Several groups or clusters can be seen, at redshifts of about $1200\,\mathrm{km\,s^{-1}}$, $2400\,\mathrm{km\,s^{-1}}$ and at $5200\,\mathrm{km\,s^{-1}}$ (associated with Perseus-Pisces). There is a high contribution from the Perseus-Pisces Cluster to the sample. The nearby group that contains Maffei I, Maffei II and IC342, can also be seen at velocities $< 200 \,\mathrm{km\,s^{-1}}$. Judging from its position in the sky and its redshift, Dwingeloo 1 is likely to be a part of this group.

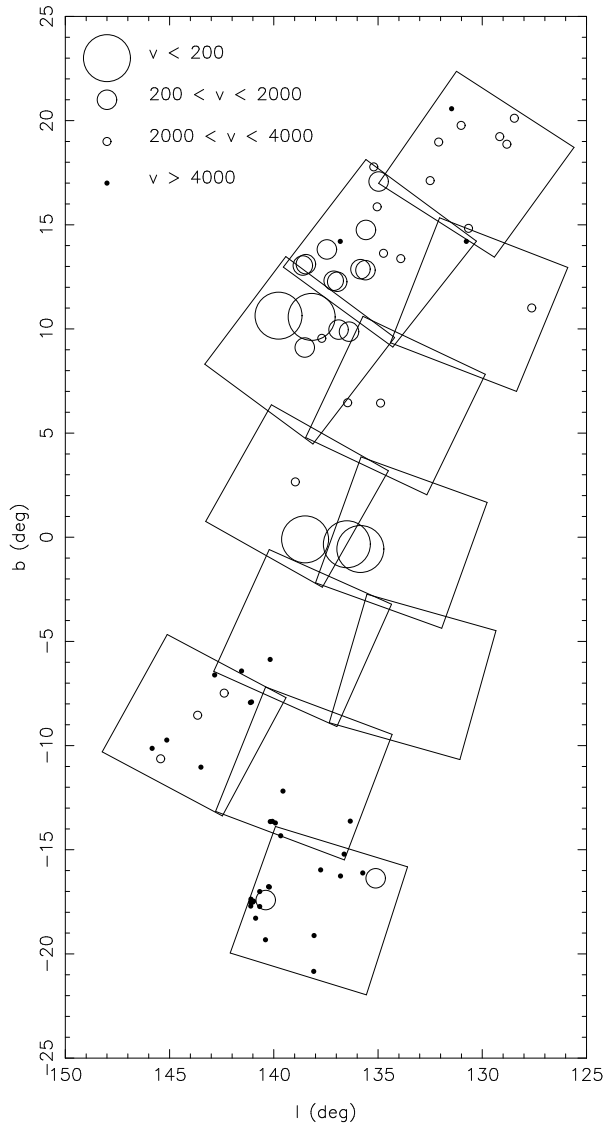

**Figure 8.** The spatial distribution of our 70 $d \geq 1'$ galaxies that have listed redshifts in Huchra's ZCAT. Dwingeloo 1 (candidate J24) at $(l, b) = (138.5, -0.1)$ is included.

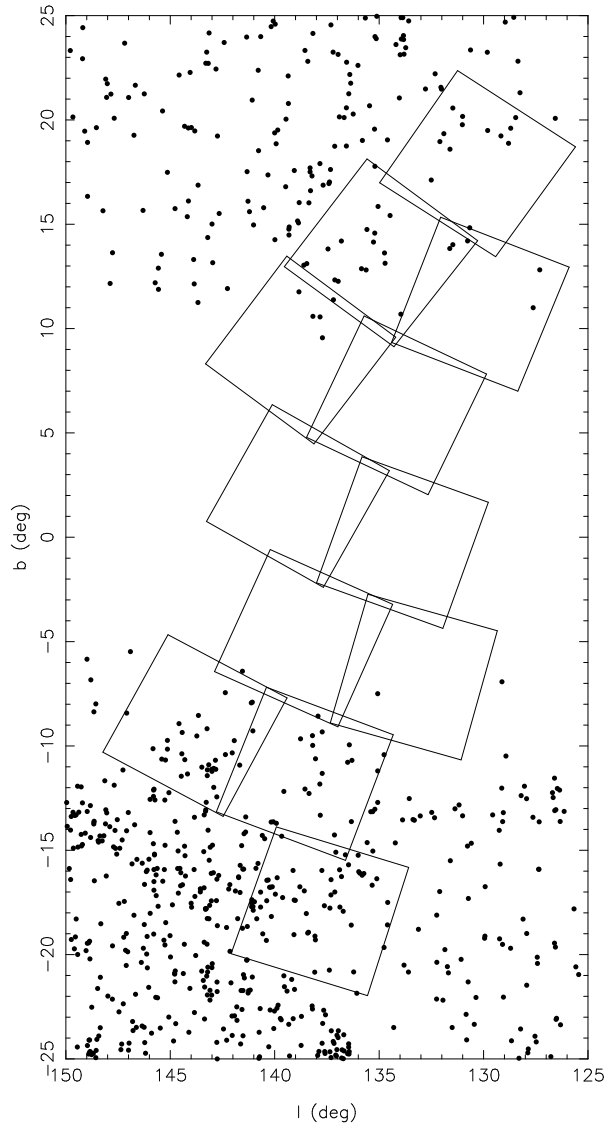

**Figure 9.** The UGC galaxies in the search region, plotted are also the plate edges.

In Fig. 8 these galaxies are plotted with different symbols for different redshifts. The IC342/Maffei/Dwingeloo1 Group, which is at low $|b|$ dominates the nearby structure. At $v > 4000 \,\mathrm{km\,s^{-1}}$, the Perseus Pices Cluster dominates, and these galaxies are localized mainly at the lower left corner of the region.

In Fig. 6 the number density of galaxies are plotted against $b$ for the raw diameter limits of $0.8'$ and $1.0'$. For a diameter limit of $1'$, the number count stays at $\sim 0.7 \,\mathrm{deg}^{-2}$ for $b \lesssim -7$ and $b \gtrsim 9$. Number densities significantly higher than the average can be seen at $b \sim -17°$.

## 4  COMPARISON WITH CATALOGUES

### 4.1  Comparison with UGC

There are 160 UGC (Nilson 1973) galaxies within the survey region, of which 156 (98%) can be associated with the eyeballed galaxies (at all diameters) within a search radius of 2 arc minutes. The UGC galaxies are plotted in Fig. 9. In the regions covered both by UGC and by our search, there is excellent agreement in the structures revealed by the two samples. On inspection of the 4 missing UGC galaxies, they are either found to be too small to be included into the catalogue, or that they are completely invisible on the red plates. On comparison of Fig. 9 with Figs. 1 & 2, it is shown that efficient visual identification of galaxies can be made at latitude as low as $|b| \sim 5°$.

The visually-determined diameters (red) and the UGC diameters (blue) are plotted in Fig. 10 in a log-log scale. There is a correlation, with RMS diameter difference of $1.07'$ for the 156 galaxies. However, if the 3 most discrepant cases are dropped (F184=NGC 891, E94K86, G143) then the RMS is $0.67'$. The average ratio for the diameters $< d/d_{\rm UGC} >$ is 0.82 (0.79 if the 3 most discrepant cases are dropped).

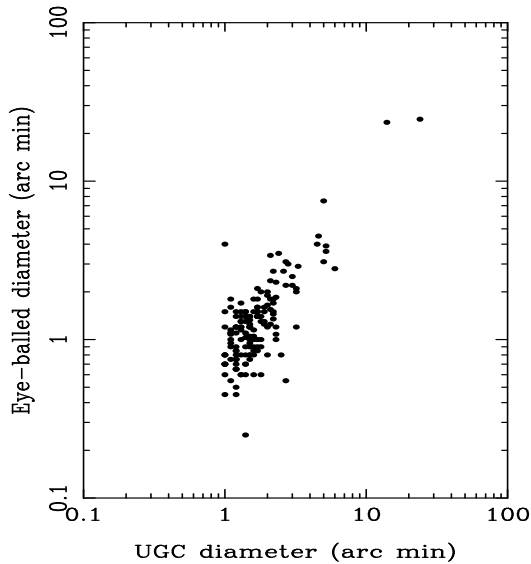

**Figure 10.** The eye-balled (red) diameters versus the UGC (blue) diameters

In order to determine the reliability of our morphological classification, in Table 3 we compare our morphological classifications with those by the UGC. There are 133 galaxies in our catalogue that has UGC morphological classification. Since the UGC morphological classification has a higher resolution than ours, to aid comparison we condense the classification in both catalogues to be either spiral, elliptical/S0 or other. In the case where there are more than one morphological classification for an object in our catalogue, we use the first entry for the comparison.

On the whole the morphological classifications for the two catalogues have reasonable agreement. The UGC ellipticals were identified quite successfully by our classification. The classification of spirals were more problematic. Firstly, about a quarter of UGC spirals were misclassified as elliptical/S0 in our catalogue, with nearly half of these attributed to early-type spirals misidentified as S0's. On the other hand, the tendency for UGC early-type galaxies to be misclassified as spirals is small. Secondly, 12% of the UGC spirals, many late-type, were misidentified as 'others' in our catalogue. The problem with mistaking early-type spirals with elliptical/S0's in our catalogue may be the consequence of working solely on red plates whilst the spiral arms are best seen in the blue; and that galaxies are morphologically classified in the UGC from their appearance on *both* the blue and red plates. Given that heavy extinction erases morphological structures such as spiral arms, we expect morphological identification to be less reliable at low $|b|$, and we recommend the morphological classification in our catalogue should be taken as a rough guide only.

**Figure 11.** The IRAS point sources under the Rowan Robinson et al. criterion. The filled circles are the 126 sources which can be associated with the our eyeballed galaxies (all diameters) within a search radius of 1 arc minute.

**Table 3.** Comparison of morphological classifications between our catalogue and the UGC.

| Our/UGC | S | E/S0 | Other |
|---------|----|------|-------|
| S       | 55 | 4    | 3     |
| E/S0    | 24 | 20   | 6     |
| Other   | 11 | 0    | 10    |

### 4.2 Comparison with IRAS

Because of the large and uniform area coverage, and low galactic extinction, many investigators have used the IRAS Point Source Catalogue for the study of Large Scale Structure by selecting sources with "galaxy colours". The success rate of picking up galaxies can be very high for high

**Figure 12.** Colour-colour plots for the IRAS point sources that fall within the plate boundaries. All points have $I_{60} \geq 0.6$ Jy and $Q_{60} \geq 2$. There are 974 sources. The sources that are identified with the visually-identified galaxies (whole sample with no diameter limit) are plotted with solid dots; there are 162 such entries. The rest of the point sources are plotted with small dots.

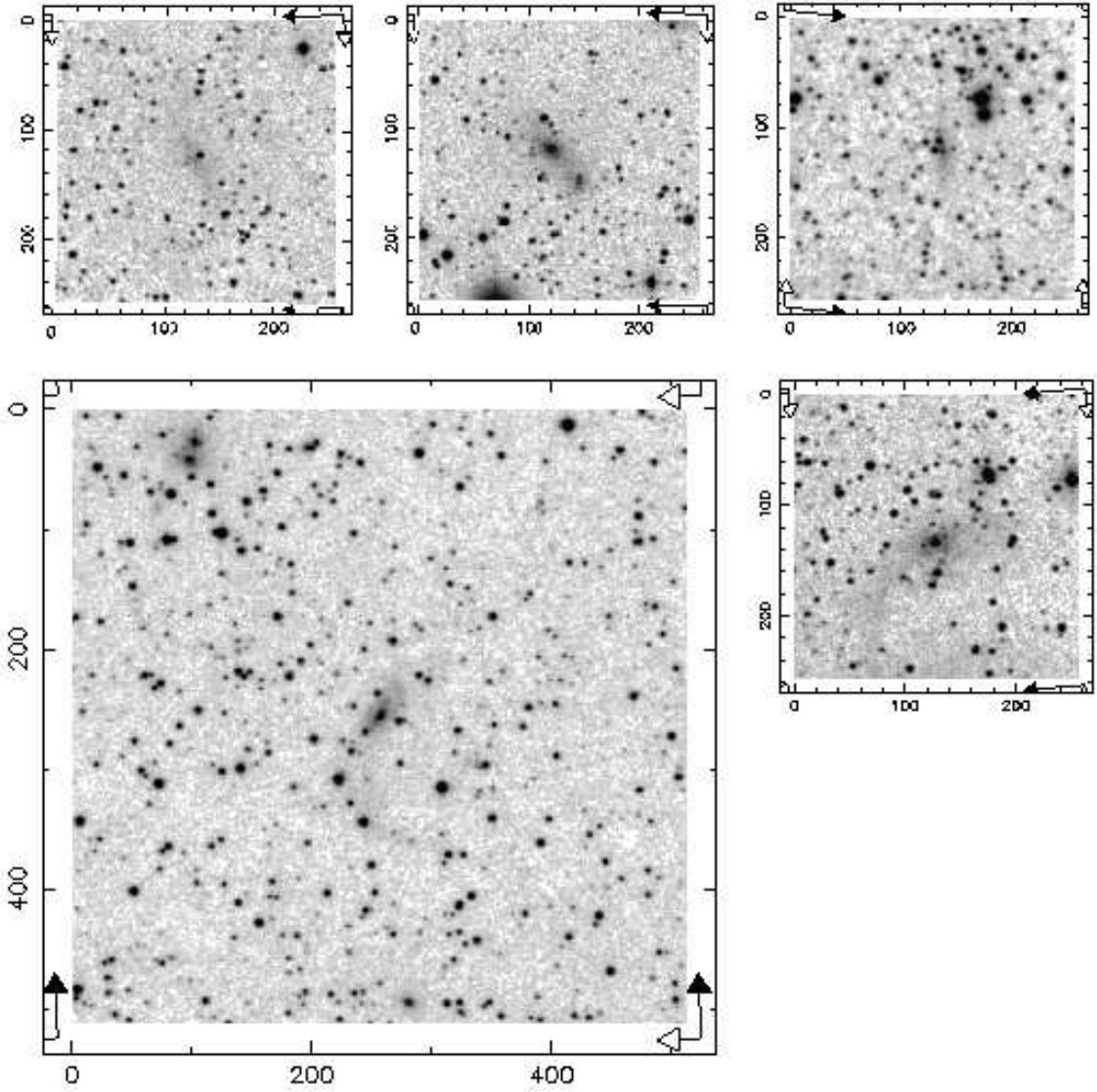

**Figure 13.** Digitized scans of selected candidates. The panels are: (clockwise from top left) J24, J17, C11J7, C14J14. E94K86. The abscissa and ordinate are in pixel number. North and east are indicated by solid and open arrows respectively.

**Table 4.** Adopted colour criteria used by various authors for our study.

| Criterion | $I_{100}$ | $Q_{100}$ | $\frac{I_{100}}{I_{60}}$ | $\frac{I_{100}}{I_{25}}$ | $\frac{I_{100}}{I_{12}}$ | $\frac{I_{60}}{I_{25}}$ | $\frac{I_{60}}{I_{12}}$ | $\frac{I_{25}}{I_{12}}$ | $\frac{I_{60}^2}{I_{12}I_{25}}$ | Addition |
|---|---|---|---|---|---|---|---|---|---|---|
| Strauss et al. | - | - | - | - | - | - | - | - | > 1 | |
| Babul & Postman | - | - | - | - | - | > 2 | - | > 0.33 | - | |
| Rowan-Robinson et al. | - | - | < 4.0 | > 0.5[1] | - | > 0.5 | > 1[2] | < 10[1] | - | |
| Yamada et al. | - | - | (0.8,5) | - | - | - | - | - | > 1 | |
| Saunders private | - | - | (1.0,4) | > 0.5[1] | - | > 2 | > 1[2] | < 10[1] | - | (a) |
| Meurs & Harmon | < 10 | ≥ 2 | (0.8,4) | - | - | (0.63,10) | - | - | - | |
| Lu et al. (proposed) | (2.5,8) | ≥ 2 | (1.13,4) | - | > 5 | > 2 | - | - | - | (b) |

All the above criteria satisfy the basic criterion: $I_{60} \geq 0.6$ Jy, $Q_{60} \geq 2$. The values in brackets are upper and lower-limits (i.e. lower-limit < quantity < upper-limit ). Unless specified, the flux ratios include those that are calculated using the upper-limit flux values.
[1] If $Q_{25} \geq 2$.
[2] If $Q_{12} \geq 2$.
[a] $CC_{60} > 0.97$; exclude sources extended at 60 $\mu$m in high source-density areas (Rowan-Robinson et al. 1991)
[b] Both $CC_{60}$ and $CC_{100} > 0.97$

Table 5. Efficiencies of various colour criteria in picking up our sample of optically identified galaxies, at a diameter limit of $0.6'$.

| Criterion | IRPS | | | OIA's ($0.6'$) | | |
|---|---|---|---|---|---|---|
| | $\|b\| > 8°$ | $\|b\| \leq 8°$ | Total | $\|b\| > 8°$ | $\|b\| \leq 8°$ | Overall |
| $I_{60} \geq 0.6$ Jy, $Q_{60} \geq 2$ (IRPSC60) | 227 | 747 | 974 | 77 (34%) | 34 (5%) | 111 (11%) |
| Strauss et al. | 189 | 660 | 849 | 77 (41%) | 34 (5%) | 111 (13%) |
| Babul & Postman | 182 | 629 | 811 | 76 (42%) | 32 (5%) | 108 (13%) |
| Rowan-Robinson et al. | 128 | 168 | 296 | 66 (52%) | 27 (16%) | 93 (31%) |
| Yamada et al. | 121 | 159 | 280 | 66 (55%) | 27 (17%) | 93 (33%) |
| Saunders private | 78 | 74 | 152 | 40 (51%) | 16 (22%) | 56 (37%) |
| Meurs & Harmon | 82 | 48 | 130 | 47 (57%) | 17 (35%) | 64 (49%) |
| Lu et al. (proposed) | 41 | 29 | 70 | 30 (73%) | 10 (34%) | 40 (57%) |

The first column lists the colour criterion applied to the IRAS point sources and columns 2, 3 and 4 list the numbers of point sources selected under this criterion at high $|b|$, low $|b|$ and in total respectively. These colour-selected IRAS point sources are then searched for optical associations (OIA's) in our sample of optically identified galaxies down to $d = 0.6'$, to test the success rate in finding optical galaxies. The numbers of OIA's for high $|b|$, low $|b|$ and overall are listed in columns 5, 6 and 7 respectively, with the percentage recovered listed in brackets.

$|b|$'s, but can fall substantially for $|b| \lesssim 5°$, presumably due to confusion with galactic infrared emission. Nevertheless, some groups have demonstrated that detection of galaxies can still be done at lower efficiencies in these regions (Lu et al. 1990, Rowan Robinson et al. 1991, Yamada et al. 1993). In this section we shall examine the efficiency in picking up the visually-identified galaxies by the various colour criteria, based on the 4 IRAS wavelengths (12, 25, 60, 100 $\mu$m).

Within the 12 plates we have scanned there are 1219 IRAS point sources with definite 60 $\mu$m fluxes $I_{60}$ (indicated by $Q_{60} \geq 2$, where $Q$ is the IRAS 'quality' flag), of which 974 have $I_{60} \geq 0.6$ Jy, the known completeness limit of IRAS. From here onwards, we shall use the 974 $Q_{60} \geq 2$, $I_{60} \geq 0.6$ Jy sources as our 'basic' IRAS sample, and refer to it as IRPSC60. Although some groups apply corrections to the IRAS fluxes in their sample (such as Strauss et al. 1990), we shall use the colour-uncorrected fluxes for the rest of the work.

Associations with the eyeballed galaxies are made by specifying a search radius of $1'$ and taking the nearest candidate as the counterpart. This works well as $\sim 2\%$ of the IRPSC60 have more than 1 optical candidate within the search radius. The number of IRPSC60 sources paired with the eyeballed galaxies are 88, 111 and 162 for $d \geq 0.8', 0.6'$ and the whole catalogue respectively. Using the more restrictive IRAS criteria of Rowan-Robinson et al. (1991) we find that 126 of his IRAS galaxies have counterparts in our entire optical sample, as shown in Fig. 11.

In Fig.12 the various colour-colour plots for the IRAS sources, with and without optical counterparts, are plotted. The majority of the paired galaxies occupy a fairly well defined region in colour space. In particular, there is a strong segregation of sources at $I_{100}/I_{60} \approx 1$, $I_{100}/I_{60} \approx 4$, $I_{60}/I_{25} \approx 2$, and at $I_{60}/I_{12} \approx 1$. The sources with $I_{100}/I_{60} \gtrsim 4$ tend to be galactic cirrus, whilst the group of sources with $I_{60}/I_{25} < 1$, $I_{60}/I_{12} < 1$ are stars (Meurs & Harmon 1988). Since a number of our optically identified candidates have $I_{100}/I_{60} > 4$, it suggests that the distinction between galaxies and galactic cirrus does not have a well defined boundary, and that some of our candidates could be galactic cirrus. Although the most efficient way of separating real galaxies from galactic cirrus is a cutoff at $I_{100}/I_{60} = 4$, such a cutoff will exclude some galaxies. When IRAS colours are used to identify galaxies, there is a conflict between maximizing the efficiency at the expense of loosing some galaxies, and maximizing the number of galaxies at the expense of including many more cirrus sources.

A comparison of different IRAS colour criteria, explained in Table 5, is provided in Table 4. For example, out of 974 IRPS60, the Yamada et al. criterion selects 280 IRAS galaxy candidates. Of those, 33 % have optical counterparts (optical-IRAS associations, hereafter OIA) in our sample down to diameter of $0.6'$.

The IRAS colour criteria can be roughly separated into three categories. The IRPSC60 under the Strauss et al. and Babul & Postman criteria recover all or nearly all of the OIA's. However, since these criteria are not very specific in confining the colour space, many galactic sources are included, thus reducing the efficiency to $\sim 13\%$ for all $b$, and $\sim 5\%$ for $|b| \leq 8°$. The selection criteria of Rowan Robinson et al. and Yamada et al. are more restrictive than the previous two, and select about 1/4 of the IRPSC60 as galaxy candidates. For these two criteria the efficiency in picking up an OIA is about $\sim 30\%$ for all $b$, and dropping to $\sim 17\%$ at low $|b|$. The Saunders criterion (private communication), which is a subset of the Rowan Robinson et al. criterion, selects $\sim 15\%$ of the IRPSC60 as galaxy candidates, but the improvement in efficiency is not obvious for this data (about 5%). The Meurs & Harmon and Lu et al. (proposed) criteria select the least number of IRPSC60 as candidates. However, their efficiencies (49 % and 57 % respectively for all $b$, and $\sim 35\%$ for low $|b|$) are higher than the other criteria. In fact the Meurs & Harmon criterion selects less IRPSC60 as candidates than the Saunders criterion, whilst recovering more of the OIA's. A major difference between these two criteria (Meurs & Harmon and Lu et al) and the rest is that these two require the fluxes to be of high quality in *both* 60 $\mu$m and 100 $\mu$m. This restriction increases the overall efficiency of galaxy identification, but at the expense of selecting fewer IRAS candidates in total.

We note with caution that since our sample is diameter limited, if an IRAS source fails to pick up an optical counterpart, it may simply be because the optical diameter falls short of the diameter limit of our sample.

## 5 DIGITIZED IMAGES OF INDIVIDUAL GALAXIES

In this section we present digitized scans of a few selected galaxies, kindly scanned by J. Pilkington, to demonstrate the appearance of galaxies under heavy extinction and the associated difficulties in distinguishing them from galactic sources (Fig. 13). The images on the Palomar plates were scanned by the PDS machine of the Royal Greenwich Observatory, with a pixel size of 16 $\mu$m, or 1.1". Recently we did a follow up imaging of these and other candidates at the 1m Wise Observatory (Israel); the results will be reported elsewhere. The comments on individual objects are:

**J24=Dwingeloo 1 (G138.52-0.11)** This galaxy has been independently discovered in a blind survey of galaxies in the ZOA by 21 cm observations (Kraan-Korteweg et al. 1994). This galaxy is under heavy extinction ($A_b \approx 5.8$, estimated by foreground HI). It is a barred spiral, in good agreement with the quoted 'SB/N' morphology, although the spiral arms are barely visible in the digitized image. At redshift of $258 \, \mathrm{km \, s^{-1}}$ relative to our Galaxy, and at an estimated distance of 3 Mpc, Dwingeloo 1 is a possible member of the Maffei/IC342 group, and may have important influence to the dynamics of nearby galaxies. The existence of this galaxy has also been confirmed by Huchtmeier et al. (1995), who named it Cass 2.

**J17 (G138.97+2.65)** This candidate has an estimated red diameter of 1.7' from the PDS scan, and an estimated foreground blue extinction of about 5.3 magnitudes. It is an IRAS galaxy and has a listed redshift of $2350 \, \mathrm{km \, s^{-1}}$ from ZCAT. There may be a small companion about 45" South-East of the core.

**C11J7 (G135.80+2.72)** Identified twice on plate overlaps, the blue extinction is estimated to be 5.7 magnitudes. It has the appearance of an edge-on spiral, with an estimated red diameter of 1.35'.

**C14J14 (G136.21+1.08)** Identified twice on the plate overlaps, it has the appearance of a spiral in the digitized image, and an estimated extent larger than that of Dwingeloo 1. However, visual examination of POSS plate reveal high galactic contamination in its vicinity, suggesting the candidate may be reflected light off galactic dust. This example shows the difficulty in distinguishing real galaxies from galactic sources in regions of high galactic contamination.

**E94K86 (G141.06-9.30)** Listed in the UGC with a blue diameter of 1', it has a relatively low foreground extinction and is one of the 3 cases with diameter very different to that listed in the UGC (§4.1). We suspect a very diffuse arm south of its nucleus with an estimated extent of 4', whilst other arm is invisible. It may have interacted with the K88K89 system approximately 5' North-East of it. Since the suspected spiral arm is very faint, the discrepancy in the quoted diameters is due to the spiral arm not being taken into account in the UGC.

## 6 DISCUSSION

We have presented a new catalogue of 2575 galaxy candidates visually examined on red Palomar plates at low Galactic latitude, where the Supergalactic Plane (SGP) is crossed by the Galactic Plane (GP), at Galactic longitude $l \sim 135°$. Comparisons with the structures seen in the IRAS and UGC catalogues indicate that galaxies can be efficiently picked up even at $|b| < 5°$.

While we are confident of the survey's completeness at diameters greater than 0.8', we recommend that the catalogue be used simply as a target list for radio, optical, and infrared observations aimed at identifying nearby galaxies (such as Dwingeloo 1). The uncertainties in the diameter estimates and the extinction corrections are sufficiently large that the galaxy surface densities are unlikely to be reliable enough for reconstructions of large-scale structure. We shall report elsewhere follow-up observations in the optical and radio of some of these candidates.

## ACKNOWLEDGMENTS

We thank Renée Kraan-Korteweg for valuable comments, discussions and encouragements. We thank Noah Brosche for carefully reading the manuscript and for his helpful comments and suggestions. We are grateful to John Pilkington of the Royal Greenwich Observatory for the digitized scans of selected candidates, and for being so helpful. Will Saunders is acknowledged for providing us information on his IRAS selection criteria. We thank the referee Chris Collins for his helpful comments. This work originated as a summer project in 1991, and GKTH thanks the Institute of Astronomy for a Summer Studentship and PPARC for a Postgraduate Research Studentship.